\newif\ifanon
\anonfalse

\documentclass[conference]{IEEEtran}
\usepackage{hyperref}
\IEEEoverridecommandlockouts
\usepackage{cite}
\usepackage{amsmath,amssymb,amsfonts}
\usepackage{algorithmic}
\usepackage{graphicx}
\usepackage{textcomp}
\usepackage{xcolor}
\usepackage{nicefrac}
\usepackage{pgfplots}
\usepackage{multirow}
\usepackage{siunitx}
\usepackage{url}
\usepackage{datetime}

\usepackage{fancyhdr}

\newcolumntype{x}[1]{>{\centering\let\newline\\\arraybackslash\hspace{0pt}}p{#1}}
\newcolumntype{y}[1]{>{\raggedright\let\newline\\\arraybackslash\hspace{0pt}}p{#1}}
\newcolumntype{z}[1]{>{\raggedleft\let\newline\\\arraybackslash\hspace{0pt}}p{#1}}

\usetikzlibrary{shapes.callouts}

\newcommand\Tstrut{\rule{0pt}{2.5ex}}	 
\newcommand\Bstrut{\rule[-1ex]{0pt}{0pt}}   

\def\BibTeX{{\rm B\kern-.05em{\sc i\kern-.025em b}\kern-.08em
    T\kern-.1667em\lower.7ex\hbox{E}\kern-.125emX}}

\definecolor{xblu}{rgb}{0.0, 0.0, 0.7}
\definecolor{xgrn}{rgb}{0.0, 0.7, 0.0}
\definecolor{xred}{rgb}{0.7, 0.0, 0.0}

\definecolor{xa1}{rgb}{0.3, 0.0, 0.8}
\definecolor{xa2}{rgb}{0.6, 0.0, 0.6}
\definecolor{xa3}{rgb}{0.9, 0.0, 0.4}

\definecolor{xb1}{rgb}{0.0, 0.6, 0.2}
\definecolor{xb2}{rgb}{0.3, 0.5, 0.3}
\definecolor{xb3}{rgb}{0.6, 0.4, 0.4}
\definecolor{xb4}{rgb}{0.9, 0.3, 0.5}

\definecolor{xc1}{rgb}{0.8, 0.9, 1.0}
\definecolor{xc2}{rgb}{0.8, 1.0, 1.0}
\definecolor{xc3}{rgb}{0.8, 1.0, 0.9}

\begin{document}

\title{Mobile Energy Requirements of the Upcoming
		NIST Post-Quantum Cryptography Standards}

\ifanon
\author{\IEEEauthorblockN{Anonymous Submission}%
	\thanks{Funding masked for anonymous submission.}
\\
IEEE Mobile Cloud 2020 \\
\\
\textit{\today}}
\else
\author{\IEEEauthorblockN{Markku-Juhani O. Saarinen}%
	\thanks{This work was supported by Innovate UK (R\&D Project Ref. 104423.)}\\
\textit{PQShield Ltd.}\\
	Oxford, United Kingdom \\
	mjos@pqshield.com	
}
\fi

\maketitle

\begin{abstract}
	Standardization of Post-Quantum Cryptography (PQC) was started by
	NIST in 2016 and has proceeded to its second elimination round.
	The upcoming standards are intended to replace (or supplement) current
	RSA and Elliptic Curve Cryptography (ECC) on all targets, including
	lightweight, embedded, and mobile systems. We present an energy
	requirement analysis based on extensive measurements of PQC candidate
	algorithms on a Cortex M4 - based reference platform. We relate
	computational (energy) costs of PQC algorithms to their data
	transmission costs which are expected to increase with new types of
	public keys and ciphertext messages. The energy, bandwidth, and latency
	needs of PQC algorithms span several orders of magnitude, which is
	substantial enough to impact battery life, user experience, and
	application protocol design. We propose metrics and guidelines for PQC
	algorithm usage in IoT and mobile systems based on our findings. Our
	evidence supports the view that fast structured-lattice PQC schemes
	are the preferred choice for cloud-connected mobile devices in most 
	use cases, even when per-bit data transmission energy cost is relatively 
	high.
\end{abstract}


\section{Introduction}

	Vulnerability of factoring-based and (elliptic curve) discrete logarithm
	cryptography (RSA, DL, ECDL, ECC) to quantum computing has been known
	since Shor published his celebrated algorithm in 1994 \cite{Sh94}.
	The following 25 years have seen several breakthroughs and a steady
	improvement in the capabilities of quantum computers -- but also the
	emergence and maturing of the field of Post-Quantum Cryptography (PQC)
	which studies public-key algorithms that are resistant to attacks by
	quantum computers in addition to classical threats.

	The first dedicated PQCrypto workshop in 2006 already featured
	talks about lattice-based, hash-based, code-based, and multivariate
	cryptography \cite{EC06}. These remain the major groups of post-quantum
	algorithms 14 years later, with the addition of the isogeny problem
	of supersingular curves \cite{AlAlAp+19,ChJoLi+16}.

	As quantum-safe ``drop-in replacements'' to RSA and ECC,
	post-quantum cryptography does not have the practical limitations of
 	Quantum Key Distribution (QKD) and other solutions that do not support
	the  public-key model and impose strict physical requirements on the
	transmission channel \cite{NC16}.

	As shown in this work, established PQC proposals include public-key
	encryption and signature algorithms that require {\it less} computational
	resources than RSA or ECC and are therefore suitable replacements in
	handheld devices, smart cards, and Internet-of-Things (IoT) applications.
	There are also PQC proposals that are are prohibitively ``heavy'' for 
	any of these targets, so understanding the individual characteristics
	of each algorithm is necessary for system design.

\subsection{Post-Quantum Standardization and Transition}
	
	In August 2015 the U.S. Committee for National Security Systems (CNSS)
	and National Security Agency (NSA) announced ``plans for transitioning 
	to quantum-resistant algorithms'' in National Security Systems (NSS),
	i.e. all governmental systems that handle classified information
	\cite{CN15,NS15}.

	Standardization bodies such as ETSI (European Telecommunications
	Standards Institute) had already initiated studies of quantum-safe
	cryptography with the view on standardizing it \cite{Peal15}, but the
	2015 CNSS/NSA post-quantum transition announcement
	created an immediate requirement for U.S. NIST (National Institute for
	Standards and Technology) to start a competitive standardization effort
	for U.S. Government use.

	NIST has studied quantum-resistant cryptography at least since 2009 
	\cite{PeCo09}. The current NIST PQC project is widely seen as a
	successor to previous, highly influential standard-selection
	competitions that led to the development of
	AES (Advanced Encryption Standard project, 1997--2001 \cite{NeBaBa+01})
	and  SHA-3 (Secure Hash Algorithm project, 2007--2012 \cite{ChPeBu+12}).

	The NIST PQC algorithm requirements were released in late 2016 and
	initial submissions were due 30 November 2017, with 69 proposals
	entering the competition. In two years the effort has proceeded to its
	second selection round with 26 proposals remaining, out of which
	17 are for encryption and key encapsulation (KEM) and 9 are intended for
	signatures \cite{AlAlAp+19,_NI19}. Table \ref{tab:nistalg} lists current
	candidates along with their rough classification and also whether
	or not they had suitable lightweight implementations for this
	study (``PQPS'').

	The NIST PQC project will go through a third
	elimination round in 2020/2021, but we already can have a fair
	understanding of the relative computational and communication
	requirements of the future standards -- it is expected that more than
	one of the current candidates will be standardized and these algorithms
	will gradually
	replace RSA and ECC in applications.  Previous experience indicates
	that such algorithm transitions typically take at least a decade to
	complete.



	\begin{table}
	\caption{Second Round NIST PQC Candidate Algorithms.}
	\label{tab:nistalg}
	\begin{center}
	\begin{tabular}{l c c c c}
	{\bf Name} 			& {\bf Type}		& {\bf Family}
						& {\bf Problem} 	& {\bf PQPS}	\\
	\hline
	BIKE				& KEM	& Codes		& QC-MDPC 		& --			\\
	Classic McEliece	& KEM	& Codes		& Goppa			& --			\\
	DILITHIUM			& Sign	& Lattice	& Fiat-Shamir	& \checkmark	\\
	Falcon				& Sign	& Lattice	& Hash\&Sign	& \checkmark	\\
	FrodoKEM			& KEM	& Lattice	& LWE			& \checkmark	\\
	GeMSS				& Sign	& Multivar.	& HFE			& --			\\
	HQC					& KEM	& Codes		& QCSD			& --			\\
	KYBER				& KEM	& Lattice	& MLWE			& \checkmark	\\
	LAC					& KEM	& Lattice 	& RLWE			& \checkmark	\\
	LEDAcrypt			& KEM	& Codes		& QC-LDPC		& --			\\
	LUOV				& Sign	& Multivar	& UOV			& --			\\
	MQDSS				& Sign	& Multivar.	& Fiat-Shamir	& --			\\
	NewHope				& KEM	& Lattice	& RLWE			& \checkmark	\\
	NTRU				& KEM	& Lattice	& NTRU			& \checkmark	\\
	NTRU Prime			& KEM	& Lattice 	& NTRU			& --			\\
	NTS-KEM				& KEM	& Codes		& Goppa			& --			\\
	Picnic				& Sign	& Symmetric	& ZKP			& --			\\
	qTESLA				& Sign	& Lattice	& Fiat-Shamir	& --			\\
	Rainbow				& Sign	& Multivar.	& UOV			& --			\\
	ROLLO				& KEM	& Codes		& Low Rank		& --			\\	
	Round5				& KEM	& Lattice 	& LWR/RLWR		& \checkmark	\\
	RQC					& KEM	& Codes		& Low Rank		& --			\\
	SABER				& KEM	& Lattice	& MLWR			& \checkmark	\\
	SIKE				& KEM	& Isogeny	& Supersingular & \checkmark	\\
	SPHINCS+			& Sign	& Symmetric	& Hash			& --			\\
	Three Bears			& KEM	& Lattice	& IMLWE			& \checkmark
	\end{tabular}
	\end{center}
	\end{table}

\subsection{NIST: PQC Key Establishment and Signatures Only}

	The 2016 NIST call for candidate algorithms \cite{NI16} limited
	the competition to three types of public-key primitives:
	Key Encapsulation Mechanisms (KEMs), public key encryption algorithms,
	and digital signature algorithms. KEMs and public key encryption
	algorithms can be trivially converted into each other -- KEM was
	originally proposed as a ``better way of doing public-key encryption''
	\cite{CrSh03}. KEMs can also be used for (ephemeral) key exchange
	and key establishment, although they may lack commutative and
	contributory properties of Diffie-Hellman required by some protocols
	\cite{UnGo18}.

	NIST defines five security levels for post-quantum algorithms; these
	correspond to best classical and quantum attacks that can be mounted
	against AES-128 (L1), SHA-256 (L2), AES-192 (L3), SHA-384 (L4), and
	AES-256 (L5). The best classical attacks against AES and SHA are
	traditional brute-force and birthday attacks so L1/L2 has 128-bit,
	L3/L4 has 192-bit, and L5 has 256-bit classical security (in addition
	to being quantum-secure). The quantum security of L1, L3, and L5
	is determined via Grover's \cite{Gr96} algorithm. For Levels L2 and L4
	quantum collision attacks \cite{ChNaSc17} can be used.

	We observe that the external functionality provided by NIST PQC
	algorithms closely matches NIST's current RSA and ECC-based
	public-key standards, namely SP 800-56 \cite{BaChRo+18,BaChRo+19}
	for key establishment and FIPS 186-4 \cite{FI13} for digital
	signatures. A far wider spectrum of quantum-secure schemes exists;
	especially hard lattice problems allows construction of virtually
	any type of scheme, from Identity- and Attribute-based Encryption
	(IBE and ABE) to Fully Homomorphic Encryption (FHE) \cite{Pe16}.

\subsection{Hybrid and Dual PQC Schemes}

	Conservative early adopters of PQC cryptography may first adopt a hybrid
	(dual, composite) approach to both key establishment and digital signatures.
	In such a scheme a PQC primitive is coupled with a conventional
	public-key algorithm for additional security assurance or interim FIPS 140
	compliance before NIST PQC algorithms are fully approved.\footnote{Dustin
	Moody (NIST), ``Revising FAQ questions on hybrid modes'', October 30, 2019.
	\url{https://groups.google.com/a/list.nist.gov/d/msg/pqc-forum/qRP63ucWIgs/rY5Sr_52AAAJ}}
	
	Newly designed systems will probably skip the hybrid stage
	as it effectively doubles the engineering effort and has a severe negative
	impact on computational overhead -- with conventional public-key
	cryptography often being the performance bottleneck.
	The main hybrid approaches are:

	\subsubsection{Hybrid Key Establishment}
	It is possible to design a hybrid key establishment method that uses a
	PQC algorithm in conjunction with a conventional method such as ECDH
	(Elliptic Curve Diffie-Hellman) in such a way that {\it both} algorithms
	(or the key derivation function that combines them) need to be attacked
	in order to compromise the overall scheme.

	\subsubsection{Dual Signatures}
	A dual signature consists of two (or more) signatures. The signatures
	can be generated by two different algorithms using two secret keys.
	The dual signature is considered valid only if both of its signatures
	are valid.

		It is possible to store a secondary (post-quantum) signature in a
	{\it non-critical} X.509 extension in a way that allows compatibility
	with existing Public Key Infrastructure (PKI) \cite{KaPaDa+18}.
	Current IETF efforts are directed towards ``composite'' dual certificates
	that do not use non-critical extensions \cite{Ou19}.

\subsection{Transport Layer Security (TLS)}

	Communication security and authentication mechanisms in IoT systems are
	currently largely based on lightweight TLS \cite{Re18} stacks running
	on top of application protocols such as MQTT \cite{BaBrBo+19,BrLa14}.
	Transition is still ongoing towards TLS protocol version 1.3 which
	was released in 2018.

	Post-quantum transition only affects the asymmetric components of
	TLS 1.3; its symmetric authenticated encryption (AEAD) and key derivation
	components (cipher suites) readily support post-quantum levels from L1
	to L5. New PQC algorithms are only used for key exchange and endpoint
	authentication, not for the transmission of bulk data streams.

	TLS is a flexible, extensible protocol that negotiates the symmetric
	cipher suites, supported key establishment methods (``groups''), and
	authentication signature algorithms as part of its client-server handshake.
	New public-key algorithms can be fairly easily incorporated in the
	framework.
	Ad hoc post-quantum TLS implementations have existed at
	least since 2015 \cite{BoCoNa+15}. The Open Quantum Safe\footnote{Open
	Quantum Safe Project: \url{https://openquantumsafe.org/}} project
	provides a free OpenSSL variant with support for a large subset of NIST
	algorithms \cite{StMo16}. In these implementations a NIST PQC KEM is
	used for key exchange and authentication is performed with a NIST PQC
	signature algorithm and certificates \cite{CrPaSt19}.

\subsection{The Cloud: Large-Scale PQC Experiments}

	The large-scale post-quantum cloud experiments ran by Amazon \cite{Ho19},
	Google \cite{La19}, and Cloudflare \cite{KwVa19} have focused on
	ephemeral hybrid key exchange methods in TLS (BoringSSL and
	S2N implementations). Algorithm selection in these experiments
	was limited to only one or two candidates, and focus was on networking
	performance; how the communication infrastructure copes
	with the changes introduced by the post-quantum transition.

	Cloudflare and Google used NTRU variant HRSS-SXY \cite{SaXaYa18},
	which has a relatively slow key generation phase and is therefore
	not ideally suited for ephemeral key exchange
	(potential key caching improves latency but not energy profile).
	They also experimented with isogeny-based proposal SIKE \cite{JaAzCa+19},
	which has the shortest messages but is computationally very expensive
	(slow). Google saw this experiment primarily as a comparison between
	isogeny-based systems and (structured) lattice schemes and tried to
	pick an ``average'' algorithm from the latter set. Google and Cloudflare
	consider HRSS as a more promising algorithm for TLS than SIKE
	\cite{KwVa19,La19}.

	The Amazon system \cite{Ho19} currently supports SIKE \cite{JaAzCa+19}
	and BIKE \cite{ArBaBe+19} algorithms. Since both of these schemes
	are relatively new and quite inefficient, we assume that their selection
	was motivated solely by research interest; Amazon's researchers are
	co-authors of both of these proposals.

	In these experiments authentication (signatures) has been seen as
	a secondary consideration to key exchange (KEMs). This is a sensible
	approach while advanced quantum computers are not available;
	authentication occurs only in the present time but a weak key
	exchange method can be attacked at any point in the future, compromising
	session confidentiality. A practical problem in trialing post-quantum
	signatures for TLS is that there is no widespread Certification Authority
	(CA) or PKI support for PQC certificates currently available.

\section{Energy Measurements}

	NIST has adopted ARM Cortex M4 CPU\cite{AR15} as their reference platform
	for lightweight PQC algorithm evaluation. Cortex M4 uses the ARMv7-m
	architecture, a somewhat stripped down version of ARM's 32-bit ISA
	(Instruction Set Architecture). Cortex M4 is licensed by ARM as an IP core
	and is often implemented as a single-chip microcontroller (MCU SoC).
	Such Cortex M4 - based MCU chips typically cost less than \$10 and are
	manufactured by NXP, ST Microelectronics, Microchip/Atmel, and many
	others. Billions of units are shipped each year, mainly for consumer
	electronics, home appliances, industrial and automotive applications.

	The PQM4\footnote{PQM4 project, source code, and results:
	\url{https://github.com/mupq/pqm4}} project has helped to port many of
	the PQC candidate algorithms to the Cortex M4 target platform. Many
	of these implementations have target-specific assembly language
	optimizations created by the design teams for this
	purpose\cite{KaRiSc+19}. We use the same implementations in our
	benchmarking but with different testing firmware and platform.
	For reference, we also measured conventional NIST-curve ECDH
	(key agreement) and ECDSA (signature) algorithms. The ECC
	implementation used is Ken MacKay's  ``micro-ecc''.\footnote{micro-ecc:
	A small and fast ECDH and ECDSA implementation for 8-bit, 32-bit, and
	64-bit processors. \url{https://github.com/kmackay/micro-ecc}}

	Note that ARM SecurCore SC300\footnote{SC300:
	\url{https://www.arm.com/products/silicon-ip-cpu/securcore/sc300}}
	secure elements are based on a closely related ARMv6-m (Cortex M3)
	architecture. STMicroelectronics alone had shipped more than a billion
	SC300-based ST33 units by early 2019.
	These chips are widely used as (e)SIMs (Subscriber Identity Modules) and
	TPMs (Trusted Platform Modules). Cortex M4 implementations
	run on SC300 targets with only minor modifications and with very similar
	cycle counts. Cortex M4 performance is, therefore, a reasonable indicator
	for expected performance on real-life SIMs, TPMs, smart cards, and other
	secure elements if no PQC-specific cryptography acceleration is 		
	implemented. However additional software side-channel countermeasures
	are appropriate on these applications since their attack model includes
	physical attacks (PQM4 implementations only have some countermeasures
	against timing attacks).

	\begin{figure}
	\begin{center}
\ifanon
	\includegraphics[width=0.45\textwidth]{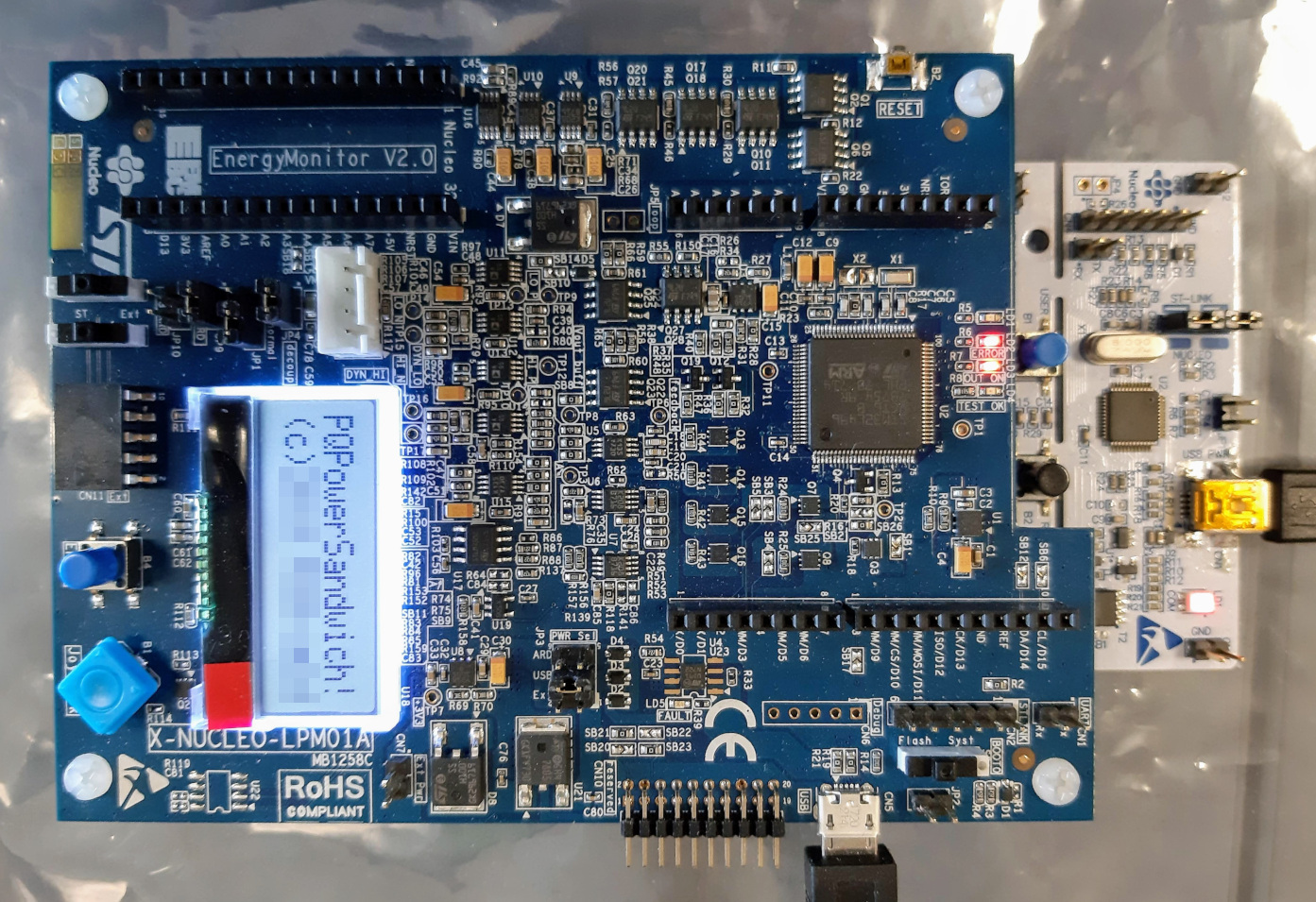}
\else
	\includegraphics[width=0.45\textwidth]{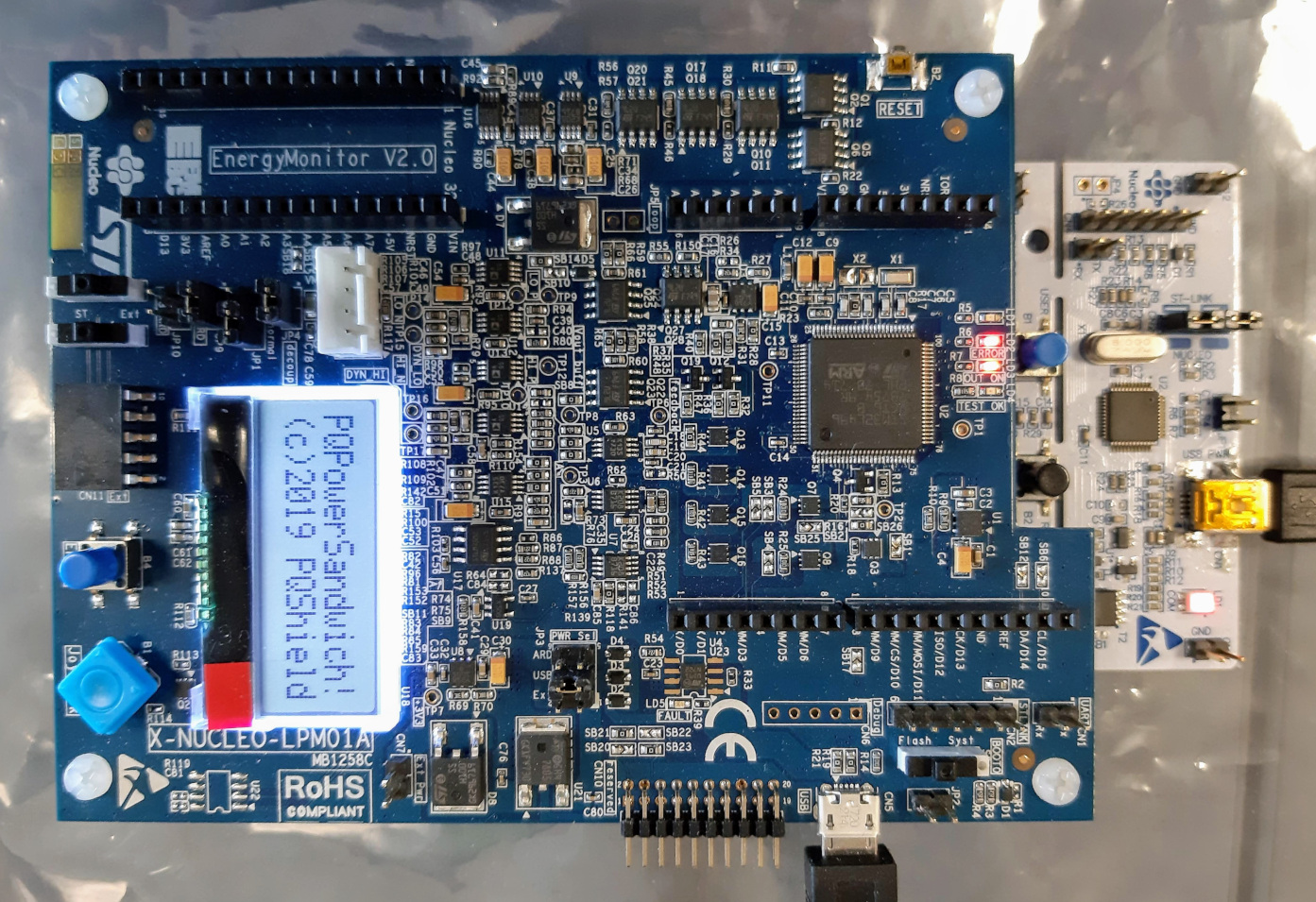}
\fi
	\end{center}
	\caption{The Post-Quantum Power Sandwich: The LPM01A ``Power Shield''
		programmable power supply installed on top of a NUCLEO-F411RE
		target.}
	\label{fig:pqps}
	\end{figure}

\subsection{PQPS, the Post-Quantum Power Sandwich}

	STMicroelectronics X-NUCLEO-LPM01A ``PowerShield'' \cite{ST18} was used
	for our power measurements. PowerShield is an
	industry-standard power and energy measurement system adopted by the
	Ultra-Low Power (ULP) group of EEMBC\footnote{EEMBC, the
	Embedded Microprocessor Benchmark Consortium develops various embedded
	system benchmarks: \url{https://www.eembc.org/}} for their
	ULPMark, IoTMark, and SecureMark benchmarks.

	The target algorithms were run on a NUCLEO-F411RE development board.
	This particular STM32F411RE \cite{ST17} Cortex M4
	MCU has a maximum clock frequency of 100 MHz, 128 kBytes of SRAM and 512
	kB of Flash memory. We chose this newer, but more limited
	board over STM32F407VG Discovery\cite{ST16} (used by the PQM4 team) as
	it supports direct, external switched-mode power supply to the
	Vcore pin of the MCU, allowing power measurements with less interference
	from peripherals. Several minor hardware modifications were necessary;
	these are documented on the project web site.

	The PowerShield measurement board is designed to be installed on the
	top of the target board; hence the name ``Post-Quantum Power
	Sandwich'' (PQPS) for this experiment. Fig. \ref{fig:pqps}
	shows the system configuration. The PowerShield was running its
	default benchmarking firmware, while the target board was running the
	firmware that we developed. Python scripts control both boards
	simultaneously and collect results on a Ubuntu Linux 18.04 host.

	We adopted most of the PQM4 \cite{KaRiSc+19} PQC implementations to
	work with our Mbed OS - based test firmware; ARM's gcc-based cross
	compilers and assemblers were used. See Table \ref{tab:nistalg} for a
	list of supported algorithms.
	We tried to add as many of the NIST PQC candidate algorithms in our
	testing suite as possible and will continue to update it.
	Many of the excluded algorithms simply do not fit on the target IoT
	platform due to excessive computational or memory requirements.
	All source code and scripts used in the PQPS experiment is open
	source and available from:
\ifanon
	{\it [masked for anonymous review]}
\else
	\url{https://github.com/mjosaarinen/pqps}.
\fi


\begin{table*}
\begin{center}
\caption{Bandwidth, CPU Cycle, Power, and Energy usage of
PQC Algorithms on Cortex M4 STM32F at 96 Mhz}
\label{tab:pqcpower}
\begin{tabular}{| c c | c c | c c c | c c c | c c c |}
\hline
	{\bf Key Establishment} & {\bf PQ}
				& \multicolumn{2}{c|}{\bf Transmit Bytes}
				& \multicolumn{3}{c|}{\bf Keypair Generation}
				& \multicolumn{3}{c|}{\bf Encapsulate}
				& \multicolumn{3}{c|}{\bf Decapsulate} \\
	Algorithm / Variant & Level
				& PubKey & CphTxt
				& $10^6$ clk & mW & Energy
				& $10^6$ clk & mW & Energy
				& $10^6$ clk & mW & Energy	\Bstrut \\
\hline
ECDH-secp256k1	& none	& 64	& 64
				& 4.108 & 65.41 & 2.799 mJ
				& 8.215 & 65.36 & 5.594 mJ
				& 4.108 & 65.45 & 2.801 mJ \\
ECDH-secp256r1	& none	& 64	& 64
				& 5.814 & 60.47 & 3.663 mJ
				& 11.63 & 59.99 & 7.267 mJ
				& 5.815 & 60.35 & 3.656 mJ \\
FrodoKEM-640-AES & [L1] & 9616	& 9720
				& 48.38 & 73.96 & 37.28 mJ
				& 47.20 & 71.34 & 35.08 mJ
				& 46.65 & 71.04 & 34.52 mJ \\
...-640-SHAKE	& [L1]	& 9616	& 9720
				& 80.01 & 79.03 & 65.87 mJ
				& 80.00 & 77.92 & 64.94 mJ
				& 79.44 & 77.84 & 64.42 mJ \\
Kyber512		& [L1]	& 800	& 736
				& 0.516 & 76.88 & 413.6 $\mu$J
				& 0.654 & 78.33 & 534.0 $\mu$J
				& 0.623 & 78.28 & 508.8 $\mu$J \\
Kyber768		& [L3]	& 1184	& 1088
				& 0.978 & 75.77 & 772.3 $\mu$J
				& 1.150 & 76.80 & 920.4 $\mu$J
				& 1.100 & 76.50 & 876.3 $\mu$J \\
Kyber1024		& [L5]	& 1568	& 1568
				& 1.575 & 77.11 & 1.265 mJ
				& 1.784 & 77.61 & 1.442 mJ
				& 1.714 & 77.47 & 1.383 mJ \\
NewHope512-CCA	& [L1]	& 928	& 1120
				& 0.591 & 78.52 & 483.9 $\mu$J
				& 0.922 & 78.09 & 750.1 $\mu$J
				& 0.906 & 77.77 & 734.0 $\mu$J \\
NewHope1024-CCA & [L5]	& 1824	& 2208
				& 1.167 & 78.91 & 959.3 $\mu$J
				& 1.785 & 78.31 & 1.456 mJ
				& 1.764 & 78.13 & 1.436 mJ \\
NTRU-HPS2048509 & [L1]	& 699	& 699
				& 78.79 & 41.94 & 34.42 mJ
				& 0.634 & 59.89 & 395.6 $\mu$J
				& 0.546 & 84.20 & 479.6 $\mu$J \\
NTRU-HPS2048677 & [L3]	& 930	& 930
				& 141.3 & 41.18 & 60.62 mJ
				& 0.943 & 59.30 & 582.7 $\mu$J
				& 0.849 & 83.52 & 739.0 $\mu$J \\
NTRU-HRSS701	& [L3]	& 1138	& 1138
				& 154.2 & 41.18 & 66.15 mJ
				& 0.398 & 79.25 & 328.8 $\mu$J
				& 0.898 & 82.91 & 776.0 $\mu$J \\
NTRU-HPS4096821 & [L5]	& 1230	& 1230
				& 212.0 & 40.90 & 90.31 mJ
				& 1.189 & 59.02 & 731.0 $\mu$J
				& 1.079 & 84.49 & 949.6 $\mu$J \\
R5ND\_1KEM\_5d	& [L1]	& 445	& 565
				& 0.342 & 73.76 & 263.5 $\mu$J
				& 0.544 & 71.82 & 407.1 $\mu$J
				& 0.729 & 69.47 & 527.9 $\mu$J \\
R5ND\_3KEM\_5d	& [L3]	& 780	& 883
				& 0.675 & 70.20 & 493.9 $\mu$J
				& 1.015 & 70.90 & 749.4 $\mu$J
				& 1.298 & 69.89 & 945.3 $\mu$J \\
R5ND\_5KEM\_5d	& [L5]	& 972	& 1095
				& 1.229 & 66.42 & 850.0 $\mu$J
				& 1.785 & 66.00 & 1.227 mJ
				& 2.339 & 64.47 & 1.571 mJ \\
R5N1\_1KEM\_0d	& [L1]	& 5214	& 5252
				& 5.577 & 55.63 & 3.232 mJ
				& 4.487 & 65.57 & 3.065 mJ
				& 5.340 & 65.61 & 3.650 mJ \\
R5N1\_3KEM\_0d	& [L3]	& 8834	& 8890
				& 8.914 & 54.53 & 5.063 mJ
				& 7.416 & 64.56 & 4.987 mJ
				& 8.445 & 65.19 & 5.735 mJ \\
R5N1\_5KEM\_0d	& [L5]	& 14264 & 14320
				& 32.82 & 60.90 & 20.82 mJ
				& 21.88 & 59.74 & 13.62 mJ
				& 25.59 & 60.37 & 16.10 mJ \\
LightSaber		& [L1]	& 672	& 736
				& 0.457 & 82.50 & 393.2 $\mu$J
				& 0.651 & 82.41 & 559.3 $\mu$J
				& 0.677 & 83.23 & 587.8 $\mu$J \\
Saber			& [L3]	& 992	& 1088
				& 0.899 & 82.32 & 771.2 $\mu$J
				& 1.170 & 82.71 & 1.008 mJ
				& 1.209 & 83.44 & 1.051 mJ \\
FireSaber		& [L5]	& 1312	& 1472
				& 1.455 & 81.87 & 1.241 mJ
				& 1.791 & 82.20 & 1.533 mJ
				& 1.854 & 82.63 & 1.596 mJ \\
LAC128			& [L1]	& 544	& 712
				& 2.275 & 49.61 & 1.176 mJ
				& 3.993 & 49.11 & 2.043 mJ
				& 6.064 & 50.58 & 3.195 mJ \\
LAC192			& [L3]	& 1056	& 1188
				& 7.546 & 49.81 & 3.916 mJ
				& 10.01 & 50.53 & 5.267 mJ
				& 18.53 & 50.23 & 9.698 mJ \\
LAC256			& [L5]	& 1056	& 1424
				& 7.686 & 50.94 & 4.079 mJ
				& 13.56 & 50.66 & 7.158 mJ
				& 22.22 & 50.59 & 11.71 mJ \\
BabyBear		& [L2]	& 804	& 917
				& 0.657 & 71.04 & 486.7 $\mu$J
				& 0.825 & 69.60 & 598.8 $\mu$J
				& 1.276 & 68.06 & 904.5 $\mu$J \\
MamaBear		& [L4]	& 1194	& 1307
				& 1.280 & 70.58 & 941.4 $\mu$J
				& 1.501 & 69.33 & 1.084 mJ
				& 2.137 & 68.21 & 1.518 mJ \\
PapaBear		& [L5]	& 1584	& 1697
				& 2.126 & 70.16 & 1.554 mJ
				& 2.400 & 69.16 & 1.729 mJ
				& 3.232 & 68.28 & 2.299 mJ \\
ntrulpr653		& [L2]	& 897	& 1025
				& 56.57 & 47.06 & 27.73 mJ
				& 112.6 & 47.07 & 55.24 mJ
				& 168.4 & 47.09 & 82.59 mJ \\
ntrulpr761		& [L3]	& 1039	& 1167
				& 76.64 & 47.74 & 38.12 mJ
				& 152.7 & 47.87 & 76.14 mJ
				& 228.3 & 47.23 & 112.3 mJ \\
ntrulpr857		& [L4]	& 1184	& 1312
				& 97.03 & 47.13 & 47.64 mJ
				& 193.3 & 47.25 & 95.16 mJ
				& 289.2 & 47.11 & 141.9 mJ \\
sntrup653		& [L2]	& 994	& 897
				& 600.3 & 46.38 & 290.1 mJ
				& 56.68 & 47.76 & 28.21 mJ
				& 171.4 & 46.69 & 83.34 mJ \\
sntrup761		& [L3]	& 1158	& 1039
				& 813.4 & 46.32 & 392.5 mJ
				& 76.76 & 47.32 & 37.84 mJ
				& 232.4 & 46.49 & 112.6 mJ \\
sntrup857		& [L4]	& 1322	& 1184
				& 1027	& 46.69 & 499.5 mJ
				& 97.16 & 48.05 & 48.63 mJ
				& 293.9 & 46.20 & 141.4 mJ \\
SIKEp434		& [L1]	& 330	& 346
				& 666.0 & 67.89 & 471.0 mJ
				& 1091	& 68.18 & 774.7 mJ
				& 1163	& 68.17 & 826.1 mJ \\
SIKEp503		& [L2]	& 378	& 402
				& 1004	& 68.69 & 718.8 mJ
				& 1656	& 68.81 & 1.187 J
				& 1761	& 69.01 & 1.266 J \\
SIKEp610		& [L3]	& 462	& 486
				& 1880	& 68.72 & 1.346 J
				& 3460	& 69.11 & 2.491 J
				& 3480	& 69.10 & 2.505 J \\
SIKEp751		& [L5]	& 564	& 596
				& 3404	& 67.76 & 2.403 J
				& 5521	& 68.40 & 3.934 J
				& 5930	& 68.59 & 4.237 J \\
\hline
\multicolumn{13}{c}{}\\
\hline
\Tstrut
	{\bf Signature} & {\bf PQ}
				& \multicolumn{2}{c|}{\bf Transmit Bytes}
				& \multicolumn{3}{c|}{\bf Keypair Generation}
				& \multicolumn{3}{c|}{\bf Sign}
				& \multicolumn{3}{c|}{\bf Verify} \\
	{Algorithm / Variant} & Level
				& PubKey & SigLen
				& $10^6$ clk & mW & Energy
				& $10^6$ clk & mW & Energy
				& $10^6$ clk & mW & Energy	\Bstrut \\
\hline
ECDSA-secp256k1 & none	& 64	& 64
				& 4.109 & 64.02 & 2.741 mJ
				& 4.475 & 64.96 & 3.028 mJ
				& 4.546 & 65.00 & 3.078 mJ \\
ECDSA-secp256r1 & none	& 64	& 64
				& 5.814 & 59.14 & 3.582 mJ
				& 6.185 & 59.97 & 3.864 mJ
				& 6.639 & 59.88 & 4.142 mJ \\
Dilithium2		& [L1]	& 1184	& 2044
				& 1.328 & 77.58 & 1.073 mJ
				& 4.663 & 77.75 & 3.777 mJ
				& 1.389 & 77.49 & 1.121 mJ \\
Dilithium3		& [L2]	& 1472	& 2701
				& 2.172 & 77.78 & 1.760 mJ
				& 7.212 & 76.48 & 5.746 mJ
				& 2.116 & 77.37 & 1.705 mJ \\
Dilithium4		& [L3]	& 1760	& 3366
				& 2.930 & 78.25 & 2.389 mJ
				& 7.263 & 77.11 & 5.834 mJ
				& 2.997 & 78.02 & 2.436 mJ \\
Falcon-512		& [L1]	& 897	& 690
				& 182.2 & 62.53 & 118.7 mJ
				& 39.57 & 55.91 & 23.05 mJ
				& 0.493 & 67.12 & 345.0 $\mu$J \\
Falcon-512-tree & [L1]	& 897	& 690
				& 200.9 & 62.31 & 130.4 mJ
				& 18.19 & 60.18 & 11.40 mJ
				& 0.492 & 66.80 & 342.7 $\mu$J	\\
Falcon-1024		& [L5]	& 1793	& 1330
				& 380.2 & 58.72 & 232.6 mJ
				& 79.36 & 54.67 & 45.19 mJ
				& 1.013 & 65.37 & 690.0 $\mu$J \\
\hline
\end{tabular}
\end{center}
\end{table*}

\subsection{Measurement Results}

	Table \ref{tab:pqcpower} summarizes our results. Some NIST-curve ECDH
	and ECDSA measurements are offered as a reference;
	For PQC algorithms the table contains the claimed post-quantum
	security level L1 $\cdots$ L5 \cite{NI16}, followed by
	public key (PubKey), ciphertext (CphTxt)
	and signature (SigLen) lengths in bytes and the actual measurements
	for keypair generation, encapsulation, decapsulation (corresponding to
	public-key encryption and decryption), and creating and verifying
	signatures. Each measurement consists of timing in millions of clock
	cycles, average power in milliwatts, and energy in Joules.

	The MCU was clocked at \SI{96}{\MHz} during the experiment.
	The PowerShield was programmed to regulate the voltage at
	\SI{3.00}{\volt} and integrate energy consumption from dynamic current
	(which the device samples with high frequency). The current varied
	between \SI{10.6}{mA} and \SI{37.8}{mA} during acquisition.
	The on-board temperature sensor reported between \SI{20.0}{\celsius}
	and  \SI{27.0}{\celsius} during the experiment, with \SI{78}{\percent}
	of measurements at ${26 \pm 1}$\SI{}{\celsius}. The temperature
	stayed within \SI{5}{\celsius} from each board self-calibration.

	Each acquisition was synchronized with the execution of the target
	algorithm at microsecond precision by using one of the interface
	pins (d7) as a trigger (a feature of PowerShield firmware). Each
	measurement ran for \SI{10000}{\ms} (roughly $10^9$ cycles) which is
	usually sufficient for hundreds of iterations; average
	Wattage was derived from this measurement. The target board itself
	performed clock cycle counts; the energy usage of each primitive is
	derived from these quantities.

	We performed at least three full runs of all algorithm measurements
	and randomized the order in each run. This required several days of
	non-stop automated testing. The three results (which themselves are
	averages of thousands of individual measurements) were verified to
	have satisfactory consistency and precision.

	For additional implementation metrics such as code size and stack
	usage, we refer the reader to the PQM4 project \cite{KaRiSc+19}
	since we used largely the same implementations.

	\paragraph{Note on SIKE Measurements and Implementations}
	Unlike other agorithms SIKE may require more than 10 seconds for a single
	operation. The overall energy consumption is extrapolated
	from the power draw during the first 10 seconds; its instruction mix
	and power draw seems to be quite homogeneous during algorithm
	execution, however. Faster implementations of SIKE for
	Cortex M4 in assembly language are reported in \cite{SeJaAz19}, but
	these implementations have not been made publicly available
	for measurement. Reported cycle counts indicate that these are still 
	about 100 times slower than lattice-based schemes at the same PQ
	security level. On slower CPUs their multi-second latency 
	can severely affect usability in addition to the energy budget.

\subsection{On PC and Server-side Energy Consumption}

	We also measured the energy consumption of PQC algorithms on PC-class
	laptop and desktop targets. This experiment instrumented the SUPERCOP 
	benchmarking platform with Intel's built-in RAPL (Running Average Power 
	Limit) energy counters and is available under the PQPS repository.\footnote{
\ifanon
	{\it [masked for anonymous review]}
\else
	SUPPERCOP: \url{https://github.com/mjosaarinen/pqps/tree/master/suppercop}
\fi
}

 	We profiled 159 variants of 20 NIST PQC algorithms on typical desktop 
	(i7-8700) and laptop (i5-8250U) systems.
	Three different components of each algorithm were measured
	separately, bringing the total number to several hundred; the
	experiment ran for many days even though the number of compiler options
	in SUPERCOP was minimized.

\begin{figure}[h]
	\begin{center}
		\includegraphics[width=0.45\textwidth]{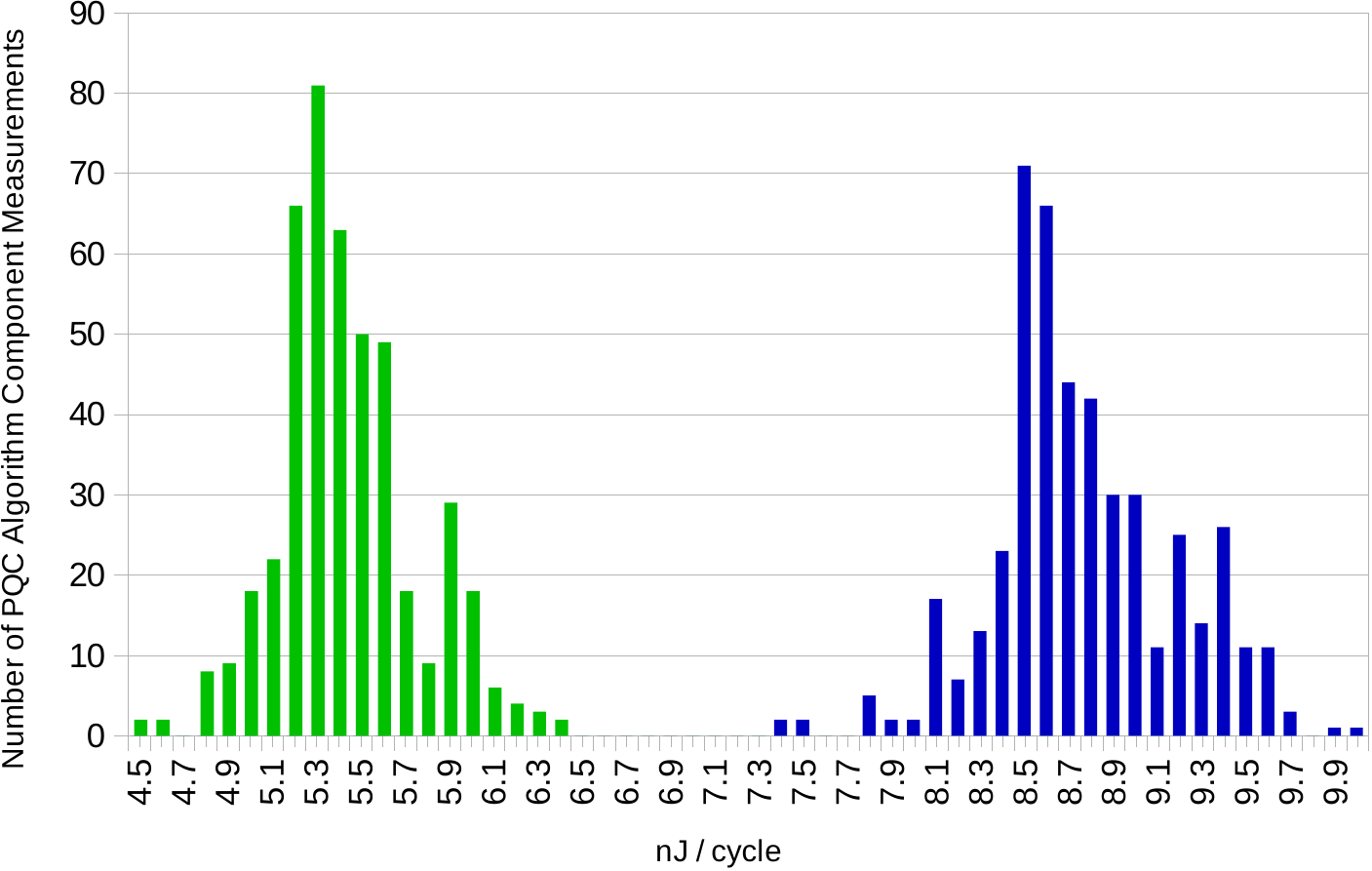}
		\caption{Running power is much less algorithm-dependent on
			PC-class chips.
		{\textcolor{xblu}{Blue:} {\it i7-8700 @ 4.6 GHz (Desktop)}}
		{\textcolor{xgrn}{Green:} {\it i5-8250U @ 3.4 GHz (Laptop)}}}
		\label{fig:rapl}
	\end{center}
	\end{figure}

	The average energy was 5.4 $\nicefrac{nJ}{\text{cycle}}$
	on the laptop system and 8.8 $\nicefrac{nJ}{\text{cycle}}$
	on the desktop system. The main observation was that power has relatively
	little variation depending on the algorithm in question on these
	platforms, being concentrated in a $\pm$ 10\% range (Fig. \ref{fig:rapl}).
	These measurements, and the ones reported in \cite{RoTaHa19}
	indicate that timing (cycle counts without related power measurements)
	leads to reasonable energy estimates on higher-end CPUs where
	power dissipation is more constant.

	\section{Analysis and Discussion}

	Most microchip technology follows the MOSFET / CMOS ``dynamic power
	equation'' \cite{HePa19} which can be written as
	\begin{equation}
		P_\text{dyn} = \alpha \cdot C \cdot V^2 \cdot f.
		\label{eqn:dynpower}
	\end{equation}
	Here $P_\text{dyn}$ = Dynamic power, $\alpha$ = activity,
	$C$ = Capacitive load, $V$ = Voltage, and $f$ = Frequency. Only the
	activity variable $\alpha$ depends on the particular
	instruction mix being executed. CPUs, of course, have also static
	power consumption; ultra-low power CPUs a lot less (by design)
	than desktop CPUs.
	
	Since frequency $f$ has a linear relationship with power (Eqn. 
	\ref{eqn:dynpower}) and cycle length $\nicefrac{1}{f}$ with algorithm 
	execution time, frequency mostly cancels out and has surprisingly little effect on
	the energy consumption of individual algorithms at CPUs ``efficiency 
	range''. However at very high 
	frequencies (flash program) memory or other components may introduce 
	additional wait states forcing algorithms to use up more clock cycles 
	for the same task. Higher clock frequencies may also require higher 
	voltage (voltage scaling); note the $V^2$ term. 

	Estimation on other targets can be performed via relative Wattage,
	cycles, and a platform scaling parameter that is often expressed in
	$\nicefrac{{\mu}A}{\text{MHz}}$. Product datasheets often contain
	vague or misleading values for this quantity; we recommend careful
	calibration with some benchmark algorithm \cite{ST13}.

	\begin{figure}[b]
	\begin{center}
	\includegraphics[width=0.47\textwidth]{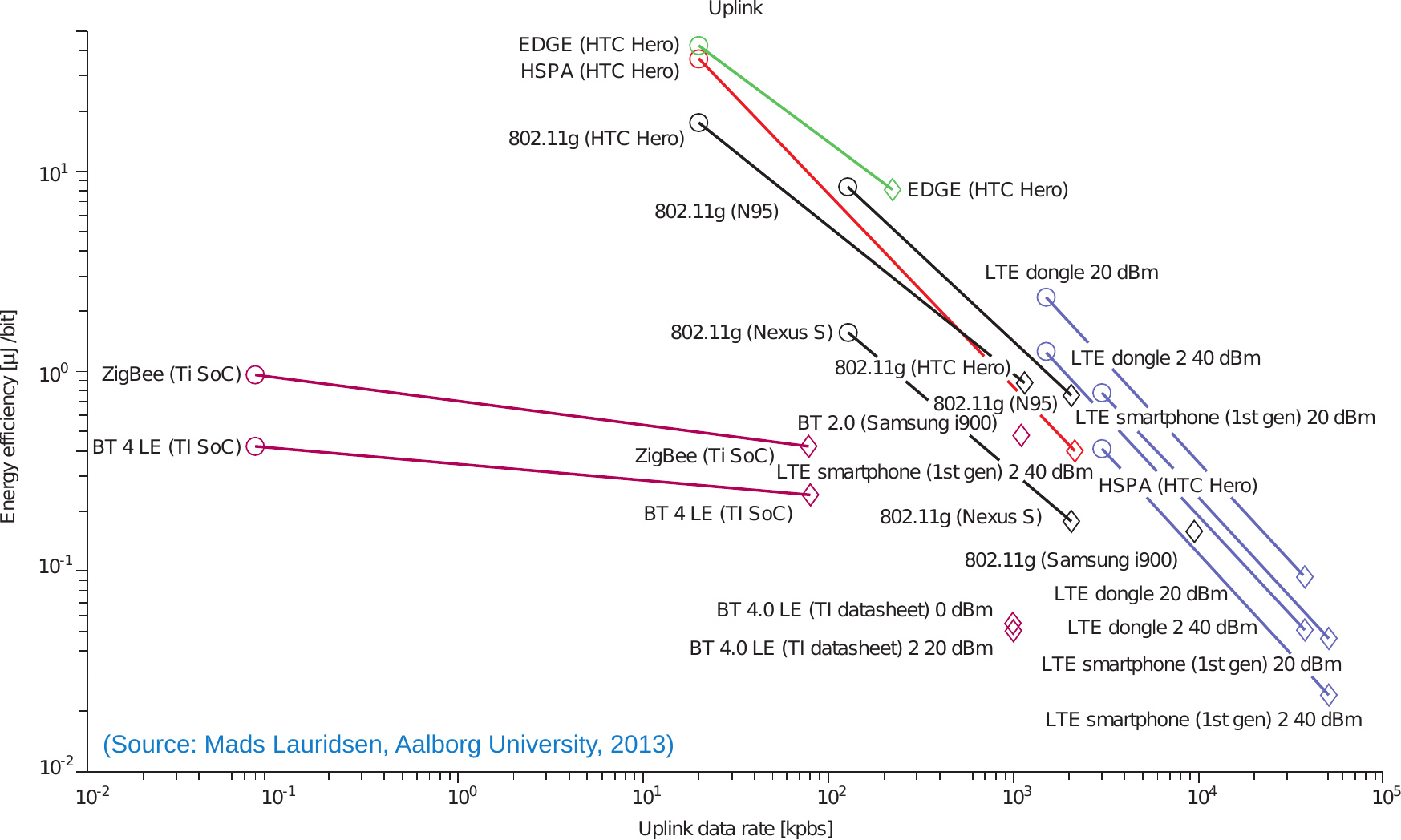}
	\end{center}
	\caption{Transmission efficiency in the pre-5G era.
		From Lauridsen et al \cite{LaNoSr+14}.}
	\label{fig:lauridsen}
	\end{figure}

	\subsection{Effect of Key and Ciphertext Lengths}

	In addition to energy consumed by computation, the length of public
	keys, signatures, and ciphertext also affect the energy consumption via
	increased radio communication. Significantly reduced transmission
	energy and time has traditionally been seen as the most important
	advantage of ECC over RSA. We can use RSA public key and ciphertext
	(or signature) lengths as a yardstick for bandwidth requirements.

	The bandwidth requirements of the PQC candidate with the shortest
	ciphertexts and public keys, Isogeny-based SIKEp434 \cite{JaAzCa+19}
	corresponds to RSA with 2600-bit keys, while the most efficient
	lattice-based proposal Round5 (R5ND\_1KEM\_5d) \cite{GaZhBh+19}
	corresponds to RSA with 4000-bit keys at the same post-quantum
	security level. However, Round5 and other structure lattice schemes
	require less than one percent of the computation time and energy of SIKE.

	We observe that the per-bit transmission energy $e_\text{xfer}$ has
	dropped from $10 \nicefrac{{\mu}\text{J}}{\text{bit}}$ during the
	2G (EDGE) era to less than $1.0 \nicefrac{{\mu}\text{J}}{\text{bit}}$
	for mobile device WiFi, HSPA, Bluetooth LE, ZigBee, and LTE	
	(See Fig. \ref{fig:lauridsen})  \cite{La14,LaNoSr+14}.
	Currently $0.1 \nicefrac{{\mu}\text{J}}{\text{bit}}$ can be
	expected with LTE or with Bluetooth LE technologies.
	Predicted 5G transmission speeds imply $e_\text{xfer}$  in
	$\nicefrac{n\text{J}}{\text{bit}}$ range.

	Simple estimation models for total energy can be used to
	choose an appropriate algorithm depending on $e_\text{xfer}$:
	\begin{center}
	\begin{tabular}{l c l l }
	$E_\text{KG}$ &+& $e_\text{xfer} | \text{pubkey} |$
			& {\it for key generation,} \\
	$E_\text{Enc}$ &+& $e_\text{xfer} | \text{ciphertext} |$
			& {\it for encapsulation,} \\
	$E_\text{Sign}$   &+& $e_\text{xfer} | \text{signature} |$
			& {\it for authentication.} \\
	\end{tabular}
	\end{center}
	where $E_\text{KG}$, $E_\text{Enc}$, and $E_\text{Sign}$ are the
	energy required for computation of keypairs, encapsulation, and
	signatures and vertical bars indicate data length in bits.

	The relevant ``energy basket'' depends on the particular
	protocol and communicating party (Alice or Bob?).
	We can also construct generalized ``index'' energy baskets
	for the purpose of comparing algorithms. For example
	\begin{equation}
		E_\text{KG}+E_\text{Enc}+E_\text{Dec} +
			e_\text{xfer}(| \text{pubkey} |+| \text{ciphertext} |)
		\label{eqn:ephemenergy}
	\end{equation}
	estimates the {\it total} energy of an ephemeral key exchange; in
	this case $e_\text{xfer}$ is the total energy to send (and receive)
	a bit.


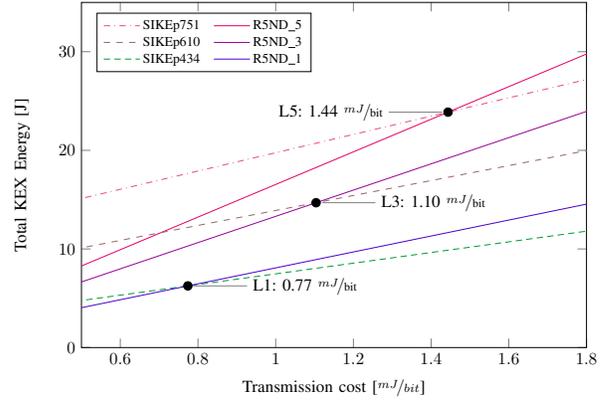
\begin{figure}
\begin{center}

\begin{tikzpicture}[thick,scale=0.8, every node/.style={scale=0.8}]

\begin{axis}[
	legend columns=2,
	legend style={nodes={scale=0.8},/tikz/column 2/.style={column sep=5pt,}},
	width=0.55\textwidth,
	height=0.31\textheight,
	legend pos=north west,
	ymin=0, ymax=35,
	xmin=0.5, xmax=1.8,
	xlabel={Transmission cost [$\nicefrac{mJ}{bit}$]},
	ylabel={Total KEX Energy [J]}
]

f5(x)=0.003647+x*8*(972+1095)
g5(x)=10.478+x*8*(564+596)


\addplot[xb4,domain=0:2, dashdotted]{10.478+x*8e-3*(564+596)};
\addlegendentry{SIKEp751}

\addplot[xa3,domain=0:2]{0.003647+x*8e-3*(972+1095)};
\addlegendentry{R5ND\_5}

\addplot[xb3,domain=0:2, dashed]{6.317+x*8e-3*(462+486)};
\addlegendentry{SIKEp610}

\addplot[xa2,domain=0:2]{0.002186+x*8e-3*(780+883)};
\addlegendentry{R5ND\_3}

\addplot[xb1,domain=0:2, densely dashed]{2.069+x*8e-3*(330+346)};
\addlegendentry{SIKEp434}

\addplot[xa1,domain=0:2]{0.001199+x*8e-3*(445+565)};
\addlegendentry{R5ND\_1}

\addplot[mark=*] coordinates {(0.7738776, 6.25413)}
	node[pin={[pin distance=7ex]0:{L1:
		0.77 $\nicefrac{mJ}{\text{bit}}$}}]{};

\addplot[mark=*] coordinates {(1.1039884, 14.68965)}
	node[pin={[pin distance=7ex]0:{L3:
		1.10 $\nicefrac{mJ}{\text{bit}}$}}]{};

\addplot[mark=*] coordinates {(1.4435437, 23.87409)}
	node[pin={[pin distance=7ex]180:{L5:
		1.44 $\nicefrac{mJ}{\text{bit}}$}}]{};

\end{axis}


\end{tikzpicture}
\caption{Energy cross-over from Round5 to SIKE is around 
	$>1 ~ \nicefrac{mJ}{\text{bit}}$.}
\label{fig:crossover}

\end{center}
\end{figure}

	\subsection{Transmission Energy Cross-over Points}

	When other relevant engineering factors are fixed and
	algorithm selection depends on transmission energy alone,
	we may use an energy model or ``basket'' such as Equation
	\ref{eqn:ephemenergy} to determine rough (order of magnitude)
	cross-over points for energy-optimal algorithm selection.

	Based on our data, the smaller message length of Isogeny-based
	PQC justifies its energy consumption over structured lattices
	when transmission energy is above the magnitude of
	$e_\text{xfer} > 1~\nicefrac{mJ}{\text{bit}}$ (Fig. \ref{fig:crossover}), 
	hundreds of times higher than $e_\text{xfer}$ of common radio interface 
	technologies used today \cite{LaNoSr+14}.\footnote{Transmission 
	energy of  $e_\text{xfer} > 1~\nicefrac{mJ}{\text{bit}}$
 	implies either continuous RF transmission power of several Watts
	(requiring a radio license) or data rate of just hundreds of bits per 
	second. Battery life is low if there is any kind of data payload;
	5 megabytes will completely drain a 40 kJ smartphone battery.}

	We may also determine the cross-over point from structured lattices
	to conventional ECC cryptography. Our data indicates ECDHE to be more 
	energy-efficient when $e_\text{xfer} > 1~\nicefrac{{\mu}J}{\text{bit}}$. 
	Actual cross-over estimates for Round5 are 1.9/2.7/3.2 
	$\nicefrac{{\mu}J}{\text{bit}}$ 
	for 256/384/512-bit EC curves (equivalent to 128/192/256-bit or L1/L3/L5
	classical security -- but no PQ security).

\section{Conclusions}

\begin{figure*}
\begin{center}

\begin{tikzpicture}[scale=0.63,remember picture, 
	note/.style={rectangle callout, font=\scriptsize,fill=#1}]

\draw[inner color=brown!30,draw=none] (4.6,3.1) ellipse (1.0 and 0.75);
\node[note=brown!30, callout absolute pointer={(4.6,3.1)}] 
	at (4.6,2.0) {ECDH-p256 (reference)};

\draw[inner color=yellow,draw=none] (2.5,5.2) ellipse (1.5 and 0.5);
\node[note=yellow, callout absolute pointer={(2.5,5.2)},anchor=west] 
	at (4.5,5.2) {Kyber and NewHope};

\draw[inner color=blue!30,draw=none] (3.0,4.1) ellipse (1.76 and 0.5);
\node[note=blue!30, callout absolute pointer={(3.0,4.1)},anchor=west] 
	at (5.0,4.1) {Three Bears};

\draw[inner color=cyan!30,draw=none] (8.5,1.1) ellipse (2.5 and 0.5);
\node[note=cyan!30, callout absolute pointer={(8.5,1.1)},anchor=south] 
	at (8.5,1.8) {NTRU Prime};

\begin{axis}[
	legend columns=2, 
	legend style={nodes=/tikz/column 2/.style={column sep=5pt,}},
	height=82mm,
	width=140mm,
	xmode=log,
	log ticks with fixed point,
	ymin=38, ymax=90,
	xmin=0.001, xmax=100,
	xlabel={Time [s] (logarithmic)},
	ylabel={Average Power [mW]}
]
\addplot[
	scatter/classes={%
		kg={mark=x,blue}, 
		sig={mark=triangle,purple},
		enc={mark=square,green}, 
		ver={mark=diamond,brown},
		dec={mark=o,red}},
	scatter, mark=x, only marks, 
	scatter src=explicit symbolic,
	nodes near coords*={\tiny\tt \alg},
	visualization depends on={value \thisrow{name} \as \alg}
] table [meta=class] {dat/kemavg1.dat};
\addlegendentry{KeyGen}
\addlegendentry{Sign}
\addlegendentry{Encaps}
\addlegendentry{Verify}
\addlegendentry{Decaps}
\end{axis}
\end{tikzpicture}%
~%
\begin{tikzpicture}[scale=0.63,remember picture, 
	note/.style={rectangle callout, font=\scriptsize,fill=#1}]

\draw[inner color=green!30,draw=none] (2.5,4.2) ellipse (1.8 and 1.1);
\node[note=green!30, callout absolute pointer={(2.5,4.2)},anchor=west] 
	at (4.5,4.2) {Round5 (structured)};

\draw[inner color=yellow,draw=none] (2.5,5.85) ellipse (1.5 and 0.5);
\node[note=yellow, callout absolute pointer={(2.5,5.85)},anchor=west] 
	at (4.5,5.85) {SABER};

\draw[inner color=blue!30,draw=none] (4.7,1.6) ellipse (2.0 and 0.5);
\node[note=blue!30, callout absolute pointer={(4.7,1.6)},anchor=south] 
	at (4.7,2.3) {LAC};

\draw[inner color=red!30,draw=none] (10.7,4.0) ellipse (1.5 and 0.5);
\node[note=red!30, callout absolute pointer={(10.7,4.0)},anchor=north] 
	at (10.7,3.3) {SIKE (SIDH)};

\begin{axis}[
	legend columns=2, 
	legend style={nodes=/tikz/column 2/.style={column sep=5pt,}},
	height=82mm,
	width=140mm,
	xmode=log,
	log ticks with fixed point,
	ymin=38, ymax=90,
	xmin=0.001, xmax=100,
	xlabel={Time [s] (logarithmic)},
	ylabel={Average Power [mW]}
]
\addplot[
	scatter/classes={%
		kg={mark=x,blue}, 
		sig={mark=triangle,purple},
		enc={mark=square,green}, 
		ver={mark=diamond,brown},
		dec={mark=o,red}},
	scatter, mark=x, only marks, 
	scatter src=explicit symbolic,
	nodes near coords*={\tiny\tt \alg},
	visualization depends on={value \thisrow{name} \as \alg}
] table [meta=class] {dat/kemavg2.dat};
\addlegendentry{KeyGen}
\addlegendentry{Sign}
\addlegendentry{Encaps}
\addlegendentry{Verify}
\addlegendentry{Decaps}
\end{axis}
\end{tikzpicture}


\begin{tikzpicture}[scale=0.63,remember picture, 
	note/.style={rectangle callout, font=\scriptsize,fill=#1}]

\draw[inner color=green!30,draw=none] (5.0,3.0) ellipse (2.0 and 1.5);
\node[note=green!30, callout absolute pointer={(5.0,3.0)}] 
	at (4.5,1.5) {Round5 (non-structured)};

\draw[inner color=blue!30,draw=none] (7.0,4.8) ellipse (1.0 and 1.0);
\node[note=blue!30, callout absolute pointer={(7.0,4.8)}] 
	at (7.0,6.0) {FrodoKEM};

\draw[inner color=yellow,draw=none] (2.0,5.75) ellipse (1.5 and 0.75);
\draw[inner color=yellow,draw=none] (2.4,2.75) ellipse (1.0 and 0.5);
\draw[inner color=yellow,draw=none] (7.8,0.5) ellipse (1.0 and 0.5);
\node[note=yellow, callout absolute pointer={(2.4,2.75)}] 
	at (2.2,4.25) {NTRU (enc,dec)};
\node[note=yellow, callout absolute pointer={(2.0,5.75)}] 
	at (2.2,4.25) {NTRU (enc,dec)};
\node[note=yellow, callout absolute pointer={(7.8,0.5)}] 
	at (7.8,1.5) {NTRU (kg)};

\begin{axis}[
	legend columns=2, 
	legend style={nodes=/tikz/column 2/.style={column sep=5pt,}},
	height=82mm,
	width=140mm,
	xmode=log,
	log ticks with fixed point,
	ymin=38, ymax=90,
	xmin=0.001, xmax=100,
	xlabel={Time [s] (logarithmic)},
	ylabel={Average Power [mW]}
]
\addplot[
	scatter/classes={%
		kg={mark=x,blue}, 
		sig={mark=triangle,purple},
		enc={mark=square,green}, 
		ver={mark=diamond,brown},
		dec={mark=o,red}},
	scatter, mark=x, only marks, 
	scatter src=explicit symbolic,
	nodes near coords*={\tiny\tt \alg},
	visualization depends on={value \thisrow{name} \as \alg}
] table [meta=class] {dat/kemavg3.dat};
\addlegendentry{KeyGen}
\addlegendentry{Sign}
\addlegendentry{Encaps}
\addlegendentry{Verify}
\addlegendentry{Decaps}
\end{axis}
\end{tikzpicture}%
~%
\begin{tikzpicture}[scale=0.63,remember picture, 
	note/.style={rectangle callout, font=\scriptsize,fill=#1}]

\draw[inner color=brown!30,draw=none] (4.4,3.1) ellipse (1.0 and 1.0);
\node[note=brown!30, callout absolute pointer={(4.3,3.1)}] 
	at (4.3,1.2) {ECDSA-p256 (ref)};

\draw[inner color=red!30,draw=none] (3.7,5.2) ellipse (2.0 and 0.5);
\node[note=red!30, callout absolute pointer={(3.7,5.2)}] 
	at (3.7,6.2) {Dilithium};

\draw[inner color=blue!30,rotate=00,draw=none] (7.7,2.7) ellipse (2.5 and 1.0);
\node[note=blue!30, callout absolute pointer={(7.7,2.7)}] 
	at (8.2,1.2) {Falcon (sign, kg)};

\draw[inner color=blue!30,rotate=00,draw=none] (2.2,3.7) ellipse (1.3 and 1.0);
\node[note=blue!30, callout absolute pointer={(2.2,3.7)}] 
	at (2.2,2.2) {Falcon (verify)};

\begin{axis}[
	legend columns=2, 
	legend style={nodes=/tikz/column 2/.style={column sep=5pt,}},
	height=82mm,
	width=140mm,
	xmode=log,
	log ticks with fixed point,
	ymin=38, ymax=90,
	xmin=0.001, xmax=100,
	xlabel={Time [s] (logarithmic)},
	ylabel={Average Power [mW]}
]
\addplot[
	scatter/classes={%
		kg={mark=x,blue}, 
		sig={mark=triangle,purple},
		enc={mark=square,green}, 
		ver={mark=diamond,brown},
		dec={mark=o,red}},
	scatter, mark=x, only marks, 
	scatter src=explicit symbolic,
	nodes near coords*={\tiny\tt \alg},
	visualization depends on={value \thisrow{name} \as \alg}
] table [meta=class] {dat/sigavg.dat};
\addlegendentry{KeyGen}
\addlegendentry{Sign}
\addlegendentry{Encaps}
\addlegendentry{Verify}
\addlegendentry{Decaps}
\end{axis}
\end{tikzpicture}

\caption{The average power of the Cortex M4 target (STM32F411RE at 96 MHz) 
varies significantly depending on the algorithm -- from 40 mW to 85 mW 
-- but computation time has a scale of four orders of magnitude.
All security levels of the same algorithm usually fit into the same cluster
since the time axis is logarithmic. Note how some algorithms such
as NTRU and Falcon have significantly more complex key generation than
other operations.}
\label{fig:avgpwr}

\end{center}

\end{figure*}
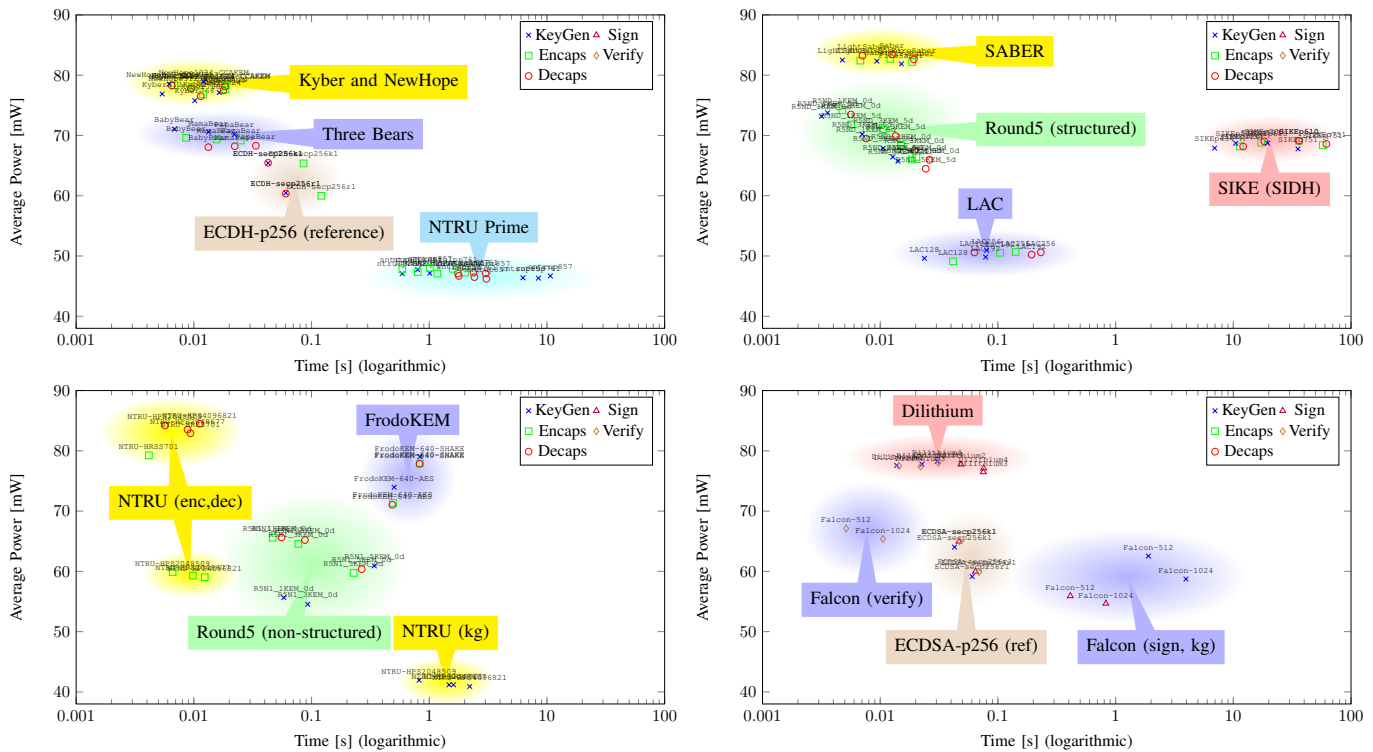

	We find that post-quantum transition can imply energy savings over
	current ECC cryptography even with current radio technologies, and 
	increasingly more with 5G's transmission speeds. Some existing ECC 
	use cases will remain that can't readily support longer messages of any 
	of the PQC alternatives (or RSA) but for TLS it is not a problem.
	Furthermore, there is a huge scale in PQC energy
	requirements (Fig. \ref{fig:avgpwr}).


	Adam Langley (Google) concludes observations from Google / Cloudflare
	post-quantum cloud experiment \cite{La19} as
	\begin{quote}
		{\it $\ldots$ Thus the overall conclusion of these experiments is that
		post-quantum confidentiality in TLS should probably be based
		on structured lattices $\ldots$}
	\end{quote}
	Our IoT experiments support this view. Here ``structured
	lattices'' refers to cryptosystems based on NTRU, RLWE, and related
	problems. The comparison was made against Isogeny-based PQC cryptography
	(SIKE) in particular.

	Kyber, NewHope, Round5, and Saber have lightweight
	variants where all three primitive operations $<$ \SI{1}{mJ}
	and public key size and ciphertext expansion is bound by 
	approximately one kilobyte (NewHope slightly more). Additionally LAC 
	and ThreeBears are very close and are faster than ECC. NTRU (HRSS) has
	a computationally expensive keypair generation which limits its 
	usability in the ephemeral (forward secure) TLS use case.
	The NTRU Prime variant is much slower than
	NTRU; not all ``structured lattice'' algorithms are equivalent.

	The Dilithium and Falcon lattice-based signature algorithms are
	suitable for mobile devices and PKI as also
	noted in \cite{KaSi19}. Falcon has smaller public keys, signatures,
	and faster signature verification so it would be preferred for
	client-side applications (i.e. when the mobile device verifies
	server certificates but does not sign). The signature performance of
	Falcon suffers from a lack of a double-precision floating point unit
	on our test platform, however, so Dilithium may be preferable when
	both parties are expected to use certificate-based authentication.

\bibliographystyle{IEEEtran}
\bibliography{pqcrypto}

\end{document}